\documentclass[pra,notitlepage,twocolumn,amsmath,amssymb,aps,superscriptaddress]{revtex4-2}

\usepackage{amsmath,amssymb,amsfonts,bbm,times}
\usepackage[dvipsnames]{xcolor}
\usepackage[T1]{fontenc}
\usepackage{braket}
\usepackage[normalem]{ulem}
\usepackage[colorlinks=true,bookmarks=false,linkcolor=RoyalBlue,urlcolor=RoyalBlue,citecolor=RoyalBlue,breaklinks]{hyperref}
\usepackage{graphicx}
\usepackage{bm}
\usepackage{physics}

\begin{document}

\title{Noise thresholds for classical simulability of non-linear Boson sampling}

\author{Gabriele Bressanini}
\affiliation{QOLS, Blackett Laboratory, Imperial College London, London SW7 2AZ, United Kingdom}
\author{Hyukjoon Kwon}
\affiliation{Korea Institute for Advanced Study, Seoul 02455, South Korea}
\author{M.S. Kim}
\affiliation{QOLS, Blackett Laboratory, Imperial College London, London SW7 2AZ, United Kingdom}
\affiliation{Korea Institute for Advanced Study, Seoul 02455, South Korea}

\begin{abstract}
Boson sampling, a computational problem conjectured to be hard to simulate on a classical machine, is a promising candidate for an experimental demonstration of quantum advantage using bosons.
However, inevitable experimental noise and imperfections, such as 
loss in the interferometer and random counts at the detectors, could challenge the sampling task from entering the regime where quantum advantage is achievable.
In this work we introduce higher order non-linearities as a mean to enhance the computational complexity of the problem and the protocol's robustness against noise, i.e., increase the noise threshold that allows to perform an efficient classical simulation of the problem.
Using a phase-space method based on the negativity volume of the relevant quasi-probability distributions, we establish a necessary non-classicality condition that any experimental proof of quantum advantage must satisfy.
Our results indicate that the addition of single-mode Kerr non-linearity at the input state preparation level, while retaining a linear-optical evolution, makes the Boson sampling protocol more robust against noise and consequently relaxes the constraints on noise parameters required to show quantum advantage.
\end{abstract}
\maketitle

\section{Introduction}
Boson sampling is a well defined computational problem, first introduced by Aaronson and Arkhipov \cite{AABS} and conjectured to be computationally hard to simulate on a classical computer, that consists in sampling from the output distribution of $N$ photons undergoing evolution via a passive linear-optical network (LON).
A passive interferometer does not contain active optical elements that alter the total photon number, i.e. the network comprises of beam splitters, phase shifters and mirrors only.
The hardness of the task stems from the fact that the transition amplitude between the initial state and the measurement outcome involves the computation of the permanent of a complex matrix \cite{scheel2004permanents}, a problem that is believed to be \#P-hard \cite{permanent_complexity}. The best known classical  algorithm for computing matrix permanents, i.e. Ryser's formula, scales exponentially with the dimension of the problem \cite{ryserpermanent}.
Under some plausible complexity-theoretic assumptions, simulating Boson sampling - even approximately - has been proven to be a classically intractable computational task and, for this reason, it is a promising candidates to experimentally show quantum advantage, i.e. the ability to outperform any classical computer on a specific task.
In fact, the advancement that photonic quantum technologies have seen in recent years \cite{Flamini_2018} made proving quantum advantage within reach with current technological capabilities.

Several variants of the original task which lie in the same complexity class have then been considered, mostly focusing on using different classes of input states such as photon-added coherent states \cite{PhotonAddedCoherentBS}, generalized cat states \cite{CatStateBS} and photon-added or photon-subtracted squeezed vacuum \cite{PhotonAddedSqueezedBS}.
Most notably, Gaussian Boson sampling (GBS) \cite{GBS} constitutes a more experimentally feasible candidate to prove quantum advantage \cite{zhong2020quantum}, as it does not require single photon generation, but rather exploits squeezed light as the initial state. Additionally, GBS finds application in graphs perfect matchings counting \cite{GBSPerfectMatchings}, measuring graph similarity \cite{GraphSimilarityGBS} and the simulation of vibronic molecular spectra \cite{VibronicSpectra_BS}.

However, inevitable noise in experimental realizations of Boson sampling might render the task classically efficiently simulable.
The effect of noise in Boson sampling and its connection to efficient classical simulability of the related computational problem have been extensively explored, considering partial photon distinguishability, losses, mode-mismatching and random counts of the detectors \cite{ClassicalSimulabilityBS_PRL,Rahimi-KeshariPRX,PartiallyDistinguishablePhotons1,PartiallyDistinguishablePhotons2,losses1,losses2,losses3,modemismatch,bulmer2021boundary}. In some cases one is able to provide sufficient conditions for efficient classical simulations of Boson sampling experiments that are expressed in form of inequalities that involve the noise parameters at play \cite{Rahimi-KeshariPRX,ClassicalSimulabilityBS_PRL}.
A possible way to make a realization of Boson sampling more robust against noise and defects is to enhance the computational complexity of the task and thus relax the constraints on noise parameters required  for and experimental demonstration of quantum advantage.
In this context, when we say that a Boson sampling protocol becomes \emph{more robust} against noise, we mean that the noise thresholds sufficient for an efficient classical simulation to be feasible do increase.

It has recently been suggested \cite{nonlinearBS} to introduce non-linear photon-photon interactions into the Boson sampling framework as a way to increase the task's complexity. In Ref.~\cite{nonlinearBS} the authors considered Fock states as input, and the non-linearity was introduced within the, otherwise linear, evolution. They provided an upper bound on complexity using a simulation method that allows to effectively induce non-linear gates using linear optical elements, auxiliary modes and photons and post-selection on photo-detection measurement outcomes.

We introduce single-mode non-linear gates in a noisy Gaussian Boson sampling problem as a way to increase its computational complexity and to relax the constraints on the maximum threshold of noise parameters necessary to prove quantum advantage. 
These higher order non-linearities are introduced at the state preparation level because, as we show in the next sections, the techniques we employ to compute these thresholds require the initial state to undergo a linear optical evolution.
To this end, we use the formalism introduced by Rahimi-Keshari in Ref.~\cite{Rahimi-KeshariPRX}, where general sufficient conditions for the efficient classical simulation of a generic quantum optics experiment - Boson sampling being a special case - are presented.
This formalism proves to be particularly helpful in studying how noise and imperfections, e.g. photon loss and sub-unit efficiency of the photo-detectors, affect the classical intractability of Boson sampling tasks.
The method is based on expressing the output probability distribution as a function of ordered phase-space quasi-probability distributions (PQDs) of the input state, the output measurements and the transition function associated with the specific quantum process. If for specific operator orderings all of these PQDs are non-negative, then an efficient classical simulation is feasible.
This result further identifies negativity as a necessary condition and as a resource to achieve quantum speed-up \cite{negativityresource1,Veitch_2012}.

Previous works studying Boson sampling protocols with non-classical input states mostly focused on proving $-$ using an array of case-dependent techniques $-$ that such tasks are at least as hard to simulate as Boson sampling \cite{PhotonAddedCoherentBS,PhotonAddedSqueezedBS,CatStateBS,GBS}.
In this paper we approach the problem from a different angle. We introduce noise in the system, in the form of loss and non-ideal detection, and gauge the enhancement in complexity due to the introduction of non-linear gates by probing an increase of the noise thresholds sufficient for an efficient classical simulation to be feasible.
Our results show how adding single-mode Kerr non-linearities at the state preparation level makes the Boson sampling task more robust to the inevitable experimental noise and imperfections that may jeopardize achieving quantum advantage.
In particular, in order to carry out analytical calculations, we consider a specific family of discrete values of the Kerr parameter that, in turn, leads to generalized squeezed cat states or superpositions of vacuum states squeezed in different directions as initial states.

The paper is structured as follows. In Section \ref{sufficientconditions} we revise some key facts about the phase-space formalism of quantum mechanics, including the concepts of characteristic functions and ordered PQDs, and outline Rahimi-Keshari's sufficient condition for an efficient classical simulation of a generic quantum optics experiment. In Section \ref{model} we introduce our model of non-linear noisy Boson sampling problem and outline the techniques used to compute noise thresholds for efficient classical simulability.
Section \ref{input1} and \ref{input2} are dedicated to investigating how two closely related families of initial states, both containing self-Kerr non-linearities, are able to increase these noise thresholds.
Lastly, in Section \ref{conclusions} we draw conclusions and give some final remarks.

\section{Sufficient conditions for efficient classical simulation of quantum optics} \label{sufficientconditions}
A generic bosonic experiment is described in terms of an $M$-mode input state $\rho_{in}$ , an $M$-mode quantum process described by a CP map $\mathcal{E}$ and a measurement on the output state $\rho_{out}=\mathcal{E}(\rho_{in})$ described by a positive operator-valued measure (POVM). The POVM elements $\lbrace \Pi_{\bm{n}} \rbrace$ satisfy the condition $\sum_{\bm{n}}\Pi_{\bm{n}}=\mathcal{I}$, where $\mathcal{I}$ is the identity operator on the $M$-mode Hilbert space.
The output probability distribution $p(\bm{n})$ of experiment is thus given by the Born rule $p(\bm{n})=\Tr{\rho_{out}\Pi_{\bm{n}}}$. In Ref.~\cite{Rahimi-KeshariPRX}, a sufficient condition for efficient classical simulability of the experiment was established, based on the well-developed theory of $\bm{s}$-ordered phase-space quasi-probability distributions ($\bm{s}$-PQD). In particular, the $\bm{s}$-PQD of a generic $M$-mode quantum state $\rho$ is defined as
\begin{equation}
    W_{\rho}^{(\bm{s})}(\bm{\beta}) = \int \frac{d^{2M}\bm{\xi}}{\pi^{2M}} \,\Phi^{(\bm{s})}_\rho (\bm{\xi}) e^{\bm{\beta}\bm{\xi}^\dagger-\bm{\xi}\bm{\beta}^\dagger} \, ,
\end{equation}
where $\Phi^{(\bm{s})}_\rho (\bm{\xi})$ is the $\bm{s}$-ordered characteristic function of $\rho$
\begin{equation}
    \Phi^{(\bm{s})}_\rho (\bm{\xi}) = \Tr{\rho D(\bm{\xi})}e^{\frac{\bm{\xi} \bm{s} \bm{\xi}^\dagger}{2}} \, .
\end{equation}
Here $\bm{s}=\text{diag}(s_1,\dots,s_M)$ is a diagonal matrix containing the $M$ ordering parameters $s_j \in\mathbb{R}$ and $D(\bm{\xi})$ is the usual $M$-mode displacement operator 
\begin{equation}
    D(\bm{\xi})=e^{\bm{\xi}\bm{a}^\dagger-\bm{a}\bm{\xi}^\dagger} \, ,
\end{equation}
$\bm{a}=(a_1,\dots,a_M)$ being the vector of annihilation operators.
The Husimi Q-function, the Wigner function and the Glauber-Sudarshan P-function are obtained for $\bm{s}=-\mathbb{I}_M$, $\bm{s}=0$ and $\bm{s}=\mathbb{I}_M$ respectively, where $\mathbb{I}_M$ denotes the $M\times M$ identity matrix.
The definition of $\bm{s}$-PQD is then straightforwardly extended to any Hermitian operator, such as the elements of a POVM.
It is worth noting that the $\bm{s}$-PQD of a Hermitian operator is a real function and that the $\bm{s}$-PQD of a quantum state is also normalized to one. 
It is then possible to express the output probability distribution of outcomes $p(\bm{n})$ in terms of quasi-probability distributions of the input state and of the POVM elements as
\begin{equation}
    p(\bm{n})=\int\! d^{2M}\bm{\beta} \int \! d^{2M}\bm{\alpha} \, \pi^M W_{\Pi_{\bm{n}}}^{(-\bm{s})}(\bm{\beta}) T_{\mathcal{E}}^{(\bm{s},\bm{t})}(\bm{\alpha},\bm{\beta}) W_{\rho_{in}}^{(\bm{t})}(\bm{\alpha}) \, .
    \label{condition_efficient}
\end{equation}
Here $W_{\Pi_{\bm{n}}}^{(-\bm{s})}$ is the $(-\bm{s})$-PQD of the POVM element $\Pi_{\bm{n}}$, $W_{\rho_{in}}^{(\bm{t})}$ is the $\bm{t}$-PQD of the input state and $T_{\mathcal{E}}^{(\bm{s},\bm{t})}$ is the transition function associated with the quantum process $\mathcal{E}$. The latter is defined as
\begin{equation}
\label{transition}
\begin{split}
    T_{\mathcal{E}}^{(\bm{s},\bm{t})}(\bm{\alpha},\bm{\beta}) & =  \int \frac{d^{2M}\bm{\zeta}}{\pi^{2M}} e^{\frac{\bm{\zeta} \bm{s} \bm{\zeta}^\dagger}{2}}e^{\bm{\beta}\bm{\zeta}^\dagger-\bm{\zeta}\bm{\beta}^\dagger}\int \frac{d^{2M}\bm{\xi}}{\pi^{2M}} e^{-\frac{\bm{\xi} \bm{t} \bm{\xi}^\dagger}{2}}\\ & e^{\bm{\xi}\bm{\alpha}^\dagger-\bm{\alpha}\bm{\xi}^\dagger} \,  \Tr{\mathcal{E}(D^\dagger (\bm{\xi}))D(\bm{\zeta})} \, .
\end{split}
\end{equation}
One can show that
\begin{equation}
    \mathcal{E}(D^\dagger(\bm{\xi}))=e^{\frac{\bm{\xi \xi}^\dagger}{2}} 
    \int \frac{d^{2M}\bm{\gamma}}{\pi^{M}} 
    e^{\bm{\gamma \xi}^\dagger - \bm{\xi \gamma}^\dagger} \mathcal{E}({\ketbra{\bm{\gamma}}}) \, .
    \label{expansion}
\end{equation}
Hence, the action of the LON on a coherent state input, i.e. $\mathcal{E}(\ketbra{\gamma})$, is everything we need in order to compute the transition function.
We are now ready to enunciate a sufficient condition for efficient classical simulation of the sampling problem outlined above.
If there exist values of $\bm{s}$ and $\bm{t}$ such that the PQD of the input, the PQD of the POVM and the transition function are all non-negative and well-behaved, i.e. they do not diverge more severely than a delta function, then a classical simulation of the sampling problem can be carried out efficiently. We point out that this formalism allows us to consider \emph{exact} simulations only, i.e. with this simulation strategy the samples are drawn according to $p(\bm{n})$ and not from  an approximation of this probability distribution.
We also stress the fact that this condition is only sufficient and, indeed, there might be other efficient simulation methods where this condition is not satisfied. 

\section{The model} \label{model}
Our model consists of a modification of GBS. The latter is a sampling problem where $M$ single-mode squeezed states are injected in a $M\times M$ linear-optical interferometer and are then measured with on/off photo-detectors at its output ports. 
The $\bm{s}$-PQD of a generic $M$-mode Gaussian state $\rho$ reads
\begin{equation}
    W_{\rho}^{(\bm{s})}(\bm{\beta})=\frac{2^M}{\pi^M}\frac{1}{\sqrt{\det{\bm{\sigma}-\bm{\tilde{s}}}}}e^{-2(\bm{\beta}-\bm{\alpha})^\intercal(\bm{\sigma}-\bm{\tilde{s}})^{-1}(\bm{\beta}-\bm{\alpha})}
    \label{sPQD_Gaussian}
\end{equation}
where $\bm{\sigma}$ and $\bm{\alpha}$ are, respectively, the covariance matrix and the vector of first moments of $\rho$ and $\bm{\tilde{s}}$ is an ordering matrix defined as
\begin{equation}
    \bm{\tilde{s}}=\bigoplus_{j=1}^M s_j \mathbb{I}_2
\end{equation}
Note that the conventions we use are such that for a single-mode coherent state $\ket{\alpha}$ the covariance matrix is the identity matrix $\sigma=\mathbb{I}_2$ and the vector of first moments reads $\bm{\alpha}=(\Re{\alpha},\Im{\alpha})$.

The $\bm{s}$-PQD of a Gaussian state is well defined and has the Gaussian form in Eq. (\ref{sPQD_Gaussian}) as long as 
\begin{equation}
    \bm{\sigma}-\bm{\tilde{s}}\geq0 \, ,
    \label{gaussian_condition}
\end{equation}
otherwise the $\bm{s}$-PQD becomes more singular delta function and does not allow for efficient sampling.
It thus follows that the $s$-PQD of a coherent state $\ket{\alpha}$ is well-behaved for $s\leq1$ and that the $s$-PQD of a squeezed vacuum $S(\xi)\ket{0}$ is properly defined for $s\leq e^{-2r}$ ($r>0$). Here the complex squeezing parameter is $\xi=re^{i\phi}$. We recall that the single-mode squeezing operator is defined as
\begin{equation}
    S(re^{i\phi})=e^{\frac{r}{2}(e^{i\phi}a^{\dagger 2}-e^{-i\phi}a^{2})} \, ,
    \label{squeezing}
\end{equation}
where $a$ and $a^\dagger$ are bosonic operators.
\\
It is well known that ideal GBS is not classically efficiently simulable \cite{GBS}. This is not necessarily true anymore if we introduce noise to the system and thus consider a realistic experimental implementation of the sampling problem.
\\
An $M$-mode passive LON is associated with an $M\times M$ transfer matrix $\bm{L}$ satisfying $\bm{L}\bm{L}^\dagger\leq\mathbb{I}$, which describes how the input modes are linearly mixed by the interferometer. For a lossless LON $\bm{L}$ is simply a unitary matrix.
Hence, a lossy LON takes an $M$-mode coherent state $\ket{\bm{\gamma}}$ to another coherent state, i.e.,
\begin{equation}
    \mathcal{E}(\ketbra{\bm{\gamma}})= \ketbra{\bm{\gamma L}} \, .
    \label{coh_state_trans}
\end{equation}
This expression stems from a simple model where we consider $M$ additional environmental modes in the vacuum state that interact with the system's actual $M$ modes via a lossless  $2M$-mode LON, whose unitary transfer matrix $\bm{{U}}$ is given by
\begin{equation}
    \bm{{U}}=
    \begin{pmatrix}
    \bm{L} & \bm{N} \\ \bm{P} & \bm{Q} 
    \end{pmatrix} \, .
\end{equation}
Eq. (\ref{coh_state_trans}) then follows from tracing out the degrees of freedom of the environment, i.e.
\begin{equation}
\begin{split}
    \mathcal{E}(\ketbra{\bm{\gamma}}) & =\Tr{\mathcal{U}\ketbra{\bm{\gamma},\bm{0} }\mathcal{U}^{\dagger}} = \\ & =\Tr{\ketbra{\bm{\gamma L} , \bm{\gamma N}  }} = \ketbra{\bm{\gamma L}}
\end{split}
\end{equation}
where $\mathcal{U}$ is the unitary operator associated with the larger $2M$-mode interferometer.
$\bm{L}$ is a submatrix of $\bm{U}$, hence the unitarity of the latter guarantees that $\bm{L}^\dagger \bm{L}\leq \mathbb{I}$.
If one further assumes that all paths in the network suffer the same amount of loss then $\bm{L}$ is simply a unitary matrix multiplied by a factor $\sqrt{\eta_L}$ with $0\leq\eta_L\leq1$.
In Appendix \ref{B} we describe how thermal noise can be added into our model and how the conclusions of this work are affected by it.

We also consider noisy on/off photo-detection characterized by sub-unit efficiency $\eta_D$ and by a random count probability $p_D$. Following Ref. \cite{Rahimi-KeshariPRX}, the POVM elements of this measurement are given by
\begin{equation}
    \Pi_0 = (1-p_D)\sum_{m=0}^\infty (1-\eta_D)^m \ketbra{m} \, ,
\end{equation}
\begin{equation}
     \Pi_1=\mathbb{I}-\Pi_0 \, 
\end{equation}
where $0\leq \eta_D \leq 1$ and $0\leq p_D \leq 1$. By noting that $\Pi_0$ is an unnormalized thermal state one obtains the following ($-s$)-PQD
\begin{equation}
    W_{\Pi_0}^{(-s)}(\bm{\beta})= \frac{1-p_D}{\pi}\cdot\frac{1}{1-\eta_D(\frac{1-s}{2})} \,\text{exp}\left[\frac{-\eta_D \vert\bm{\beta}\vert^2}{1-\eta_D(\frac{1-s}{2})} \right] \, ,
    \end{equation}
which is non-negative $-$ and properly defined $-$ for
\newline
$s\geq 1-\frac{2}{\eta_D}$. Since $\Pi_0 + \Pi_1 = \mathbb{I}$ we also have that
\begin{equation}
    W_{\Pi_1}^{(-s)}(\bm{\beta}) = \frac{1}{\pi}-W_{\Pi_0}^{(-s)}(\bm{\beta}) \, .
\end{equation}
One then easily proves that $ W_{\Pi_1}^{(-s)}$ is non-negative for 
\begin{equation}
    s\geq 1-\frac{2 p_D}{\eta_D} \equiv \overline{s} \, .
\end{equation}
Hence, the noisy photo-detection ($-s$)-PQD is non-negative for $s\geq\overline{s}$. If we then consider $M$ identical photo-detection measurements at the end of our LON, the ($-\bm{s}$)-PQD of the measurement is just the product of the ($-s_j$)-PQD of the single-mode measurements, i.e.
\begin{equation}
    W_{\Pi_{\bm{n}}}^{(-\bm{s})}=\Pi_{k=1}^M W_{\Pi_{n_k}}^{(-s_k)} \, .
\end{equation}
Consequently, the total ($-\bm{s}$)-PQD is non-negative for $s_k \geq\overline{s}\quad \forall k$.
\\
\\
The last thing that we need to consider is the transition function $T_{\mathcal{E}}^{(\bm{s},\bm{t})}$ associated with a LON described by the transfer matrix $\bm{L}$.

In Ref. \cite{Rahimi-KeshariPRX}, Rahimi-Keshari proved that it has the form of a multi-variate Gaussian function, hence non-negative and well-behaved, if and only if
\begin{equation}
    \mathbb{I}_M-\bm{L}^{\dagger}\bm{L}-\bm{s}+\bm{L}^\dagger \bm{t} \bm{L} \geq 0 \, .
    \label{trans_condition}
\end{equation}
In Appendix \ref{B} we show how this inequality is modified once thermal effects are taken into account.
\\
If the input state $\bm{t}$-PQD is non-negative for $\bm{t}\leq\overline{\bm{t}}$ and the $\bm{(-s)}$-PQD of the measurement is non-negative for $\bm{s}\geq\overline{\bm{s}}$, then Eq. (\ref{trans_condition}) is satisfied if and only if 
\begin{equation}
    \mathbb{I}_M-\bm{L}^{\dagger}\bm{L}-\overline{\bm{s}}+\bm{L}^\dagger \overline{\bm{t}} \bm{L} \geq 0 \, .
\end{equation}
If we further consider a lossy LON described by the transfer matrix $\bm{L}=\sqrt{\eta_L}\bm{U}$ and identical noisy detection at each output port as outlined above, i.e. $\overline{\bm{s}}=\overline{s}\mathbb{I}=(1-\frac{2 p_D}{\eta_D})\mathbb{I}$, it is then possible to recast the previous condition as
\begin{equation}
    \left(\frac{2 p_D}{\eta_D}-\eta_L\right)\mathbb{I}_M+\eta_L \overline{\bm{t}}\geq 0 \, .
    \label{noise_threshold}
\end{equation}
We can now compute $\overline{\bm{t}}$ for different input states and use the previous inequality to compute noise thresholds sufficient for classical simulability.
\\
\\
As a first example we might consider input coherent states, i.e. $\overline{\bm{t}}=\mathbb{I}_M$. As expected, inequality Eq. (\ref{noise_threshold}) tells us that such sampling problem is efficiently classically simulable even in the absence on noise, as this problem is equivalent to sampling from an $M$-mode coherent state. 
On the other hand, if we consider $M$ single-mode squeezed vacuum states as input, i.e. $\bigotimes_{j=1}^M S(r)\ket{0}$, then Eq. (\ref{gaussian_condition}) implies that the input state $\bm{t}$-PQD is well defined and non-negative for $\bm{t}<\overline{\bm{t}}=e^{-2r}\mathbb{I}_M$.
Hence, in this scenario, the sampling problem can be simulated efficiently if the noise parameters satisfy 
\begin{equation}
    \frac{p_D}{\eta_D}\geq \frac{\eta_L}{2}(1-e^{-2r}) \, .
    \label{noisyGBS}
\end{equation}
We stress, once again, that these noise thresholds for efficient classical simulation provide a sufficient condition only.
We also point out that Eq. (\ref{noisyGBS}) is consistent with the condition for classical simulability of noisy Gaussian Boson sampling obtained in Ref. \cite{ClassicalSimulabilityBS_PRL}. In that work the authors proved that a sufficient condition for the existence of an efficient classical simulation of a noisy GBS experiment as described above, up to error $\varepsilon$, is given by
\begin{equation}
    \sech{\left(\frac{1}{2}\Theta\left[ \ln{\left(\frac{1-2q_D}{\eta_L e^{-2r}+1-\eta_L}\right)} \right]\right)}>e^{-\varepsilon^2/4M}
\end{equation}
where $q_D=\frac{p_D}{\eta_D}$ and $\Theta(x)=\max{(x,0)}$ is the ramp function.
\\
\\
We now aim to tackle the following question.
How do these noise thresholds for efficient classical simulation change when higher order non-linearities are introduced in the model? Answering this question will tell us if, with the addition of higher order non-linearities, we can afford to allow more noise in an hypothetical experimental setup, but still have a sampling problem that is not efficiently classically simulable.
There is obviously a lot of freedom in how to introduce non-linearities in a Boson sampling protocol. In fact, they can be added to the input state preparation stage, within the evolution or as part of the measurement.
However, it is clear that if we want to apply condition Eq. (\ref{noise_threshold}) in this new setting, we still need to retain a linear optical interferometer. We will thus add the non-linear operations at the input state preparation level.

As a simple example of this paradigm, we consider single-mode Kerr non-linearities, i.e. $U(\chi)=e^{-i\chi a^{\dagger 2} a^2}$, where $\chi\in\mathbb{R}$ is the Kerr parameter.
Using Baker-Campbell-Hausdorff identities one easily displays the action of $U(\chi)$ on the annihilation operator $a$ (we provide the proof in Appendix \ref{C}), namely
\begin{equation}
    U^\dagger (\chi)a U(\chi)=e^{-2i\chi a^\dagger a}a \, .
    \label{kerr_transformation}
\end{equation}
Eq. (\ref{kerr_transformation}) shows that the Kerr transformation is an energy-dependent phase rotation of the mode. 
In the following sections we consider two closely-related classes of initial states: $S(r)U(\chi)\ket{\alpha}$ and $U(\chi)S(r)\ket{0}$.

\section{Input state $S(r)U(\chi)\ket{\alpha}$} \label{input1}
\begin{figure}[t]
\includegraphics[width=0.48\textwidth]{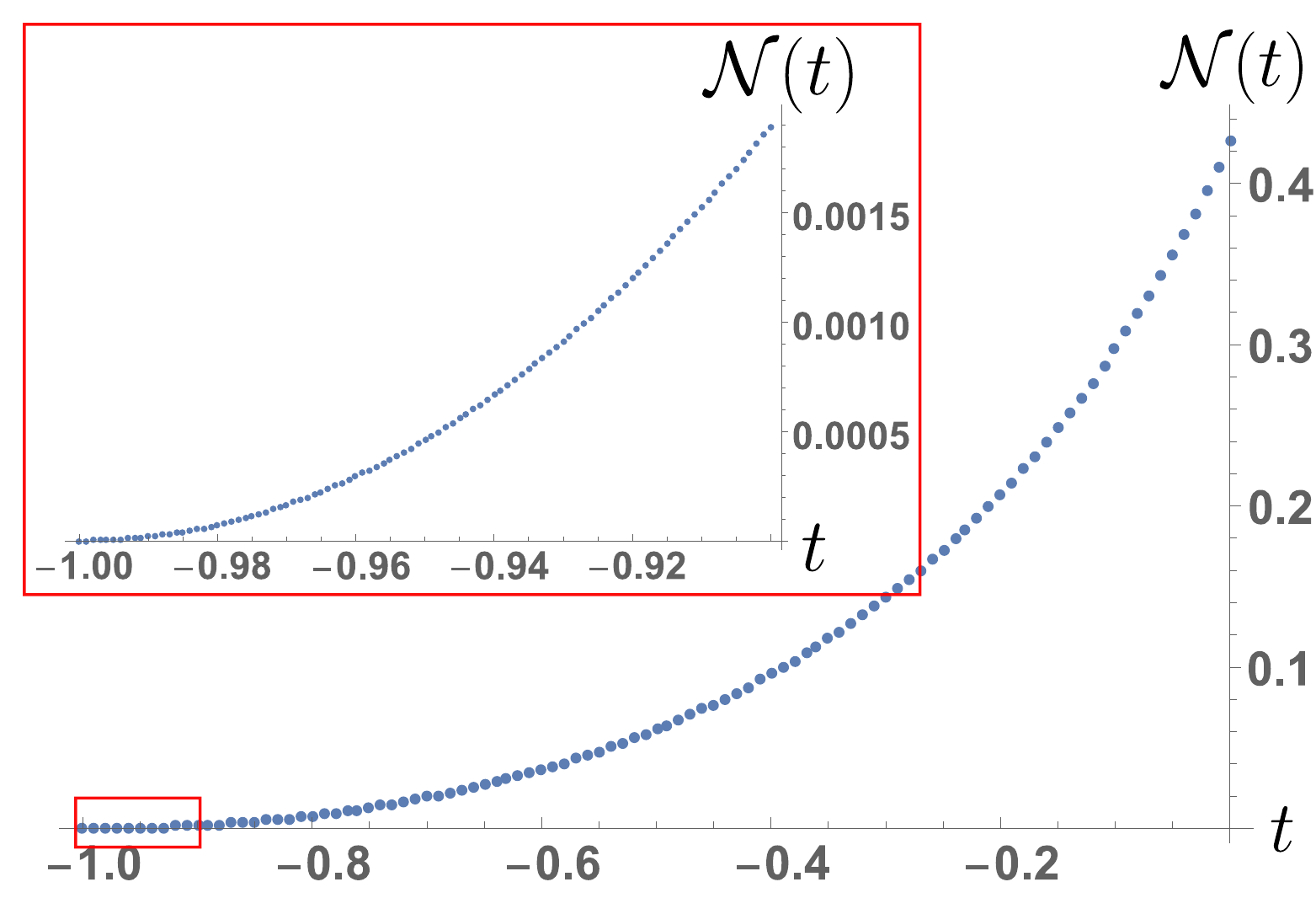}
\caption{\label{input1plot} The plot shows the negativity volume $\mathcal{N}(t)$ of the $t$-PQD associated with the input state $S(r=0.2)U(\chi=\pi/3)\ket{\alpha=1}$. 
As can be seen from the inset, the function reaches zero for $t=-1$.
Similar plots and behaviour, i.e. negativity volume approaching zero monotonically at $t=-1$, are obtained for every $r>0$, $\alpha\in\mathbb{C}$ and for every integer $m>1$.}
\end{figure}

Let us consider a sampling problem as the one described in the previous section, where the $M$-mode input state is now given by $M$ copies of $S(r)U(\chi)\ket{\alpha}$. 
The strategy to obtain the noise thresholds for efficient classical simulability remains the same: compute the $\bm{t}$-PQD of the initial state, find the value $\overline{\bm{t}}$ for which the function is non-negative for every $\bm{t}\leq \overline{\bm{t}}$ and finally use Eq.  (\ref{noise_threshold}) to compute the desired threshold. 
We remind the reader that since the input state is a tensor product of identical states, we have that $\overline{\bm{t}}=\overline{t}\mathbb{I}_M$. 
Unfortunately, one soon realizes it is not possible to obtain an analytical, easy-to-use, closed formula of the $t$-PQD of $S(r)U(\chi)\ket{\alpha}$ for a generic value of $\chi$.
A way around this is to consider discrete values of the Kerr parameter, specifically $\chi=\frac{\pi}{m}$ with $m$ integer. Notice how this corresponds to discrete-time Kerr-type interactions. In this scenario, $U(\chi)\ket{\alpha}$ produces a superposition of coherent states (cat-like states) that lie on a circumference in the phase space \cite{ProdCatKerr}. In fact, the operator
\begin{equation}
    U(\chi=\pi/m)\equiv U^{(m)}=e^{-i \frac{\pi}{m}\hat{n}(\hat{n}-1)}
\end{equation}
has nice periodic properties that allow for a useful Fourier representation of the operator which, in turn, leads us to
\begin{equation}
    \ket{\psi_m}\equiv U^{(m)}\ket{\alpha} = \sum_{q=0}^{m-1}f_q^{(o)}\ket{\alpha e^{-\frac{2\pi i q}{m}}} \, , \quad m=\text{odd}
\end{equation}

\begin{equation}
    \ket{\psi_m}\equiv U^{(m)}\ket{\alpha} = \sum_{q=0}^{m-1}f_q^{(e)}\ket{\alpha e^{-\frac{2\pi i q}{m}+\frac{i\pi}{m}}} \, , \quad m=\text{even} \, .
\end{equation}
The coefficients that appear in the sums are given by
\begin{equation}
    f_q^{(o)}=\frac{1}{m}\sum_{k=0}^{m-1}e^{\frac{2 \pi i q}{m}k}e^{-\frac{i \pi }{m}k(k-1)}
    \label{coeff1}
\end{equation}
\begin{equation}
    f_q^{(e)}=\frac{1}{m}\sum_{k=0}^{m-1}e^{\frac{2 \pi i q}{m}k}e^{-\frac{i \pi }{m}k^2}
    \label{coeff2}
\end{equation}
For example, with $m=2$ one obtains the well known Yurke-Stoler cat state up to a phase shift, namely
\begin{equation}
    \ket{\psi_2}=\frac{1}{\sqrt{2}}\ket{i\alpha}+\frac{i}{\sqrt{2}}\ket{-i\alpha} \, .
\end{equation}
We point out that in Ref. \cite{CatStateBS} evidence is presented that Boson sampling using arbitrary superpositions of coherent states as input is likely to implement a classically hard problem.
It is also worth noting that for $m=1$, i.e. $\chi=\pi$, we have that $U(\chi=\pi)=e^{-i\pi\hat{n}(\hat{n}-1)}=\mathcal{I}$. Hence, unlike squeezing, Kerr non-linearity does not always produce a non-classical effect on a classical initial state.
\\
\\
Hence, our input state is a superposition of squeezed coherent states
\begin{equation}
    S(r)U^{(m)}\ket{\alpha} = \sum_{q=0}^{m-1}f_q^{(o)}S(r)\ket{\alpha e^{-\frac{2\pi i q}{m}}} \, , \quad m=\text{odd}
\end{equation}

\begin{equation}
    S(r)U^{(m)}\ket{\alpha} = \sum_{q=0}^{m-1}f_q^{(e)}S(r)\ket{\alpha e^{-\frac{2\pi i q}{m}+\frac{i\pi}{m}}} \, , \quad m=\text{even} \, ,
\end{equation}
\\
\\
Consequently, the $t$-PQD of the input state $S(r)U(\chi)\ket{\alpha}$ is readily obtained once we have the $t$-PQD of $S(r)\ket{\alpha}\!\!\bra{\gamma}S^\dagger(r)$. 
One can prove that the $t$-ordered characteristic function of this operator reads
\begin{equation}
\begin{split}
    & \phi^{(t)}(\xi)  =\Tr{S(r)\ket{\alpha}\!\!\bra{\gamma}S^\dagger (r)D(\xi)}e^{\frac{t}{2}\vert\xi\vert^2}=
    \\ & =e^{\frac{1}{2}\left( -\vert\xi\mu-\xi^*\nu+\alpha-\gamma\vert^2  +(\gamma^*(\xi\mu-\xi^*\nu+\alpha)+\alpha^*(\xi\mu-\xi^*\nu)-c.c.)\right)}e^{\frac{t}{2}\vert\xi\vert^2}
    \label{charac1}
\end{split}
\end{equation}
where $\mu=\cosh(r)$ and $\nu=\sinh(r)$. In order to obtain Eq. (\ref{charac1}) one has to use $S^\dagger(r) D(\xi)S(r)= D(\xi\mu-\xi^*\nu)$ and the well known composition rule of consequent displacement operators, i.e. 
\begin{equation}
    D(\alpha)D(\beta)=D(\alpha+\beta)e^{\frac{1}{2}(\alpha\beta^*-\alpha^*\beta)} \, .
\end{equation}
We can then Fourier-transform the characteristic function and obtain an analytical expression for the $t$-PQD of $S(r)\ket{\alpha}\!\!\bra{\gamma}S^\dagger (r)$ and, in turn, the $t$-PQD of our input state. 
The last step to obtain the desired noise thresholds consists in finding the value $\overline{t}$ for which the $t$-PQD of the initial state is non-negative for all $t\leq \overline{t}$.
This is achieved by numerically computing the \emph{volume of negativity} $\mathcal{N}$ of the $t$-PQD as a function of the ordering parameter $t$, i.e.
\begin{equation}
  \mathcal{N}(t)= \int d^2 \bm{\beta} \, \vert W^{(t)}(\bm{\beta}) \vert - 1 \, .
\end{equation}
We have strong numerical evidence that the $t$-PQD of $S(r)U^{(m)}\ket{\alpha}$ becomes non-negative for $t\leq \overline{t}=-1\quad \forall \alpha\in\mathbb{C},\,\,\forall r>0$ and $m\geq 2$.
Figure \ref{input1plot} displays, with a specific example, the typical features of the negativity volume associated with the input state $S(r)U(\chi=\pi/m)\ket{\alpha}$ $t$-PQD. 
Finally, using $\overline{t}=-1$ in Eq. (\ref{noise_threshold}) yields the sufficient condition for efficient classical simulability
\begin{equation}
    \frac{p_D}{\eta_D}\geq \eta_L \, .
    \label{result}
\end{equation}
Recall that, without Kerr non-linearity, the threshold was $\frac{p_D}{\eta_D}\geq \frac{\eta_L}{2}(1-e^{-2r})$. Hence, more noise is needed to simulate the non-linear system, which suggests that the Kerr non-linearity does indeed increases the complexity of the sampling problem. In Appendix \ref{B} we discuss how this result is affected once we consider finite-temperature effects.
Note that the $t$-PQD with $t=-1$, i.e. the Husimi Q function, is non-negative by definition for \emph{every} state $\rho$. In fact, one can show that 
\begin{equation}
    W_{\rho}^{(-1)}(\beta)\equiv Q_\rho (\beta)=\frac{1}{\pi}\bra{\beta}\rho\ket{\beta} \, .
\end{equation}
This means that a sampling experiment as the one described above $-$ i.e. lossy LON and noisy detection $-$ is actually classically efficiently simulable for every input state if $\frac{p_D}{\eta_D}\geq \eta_L$.
Hence, we have proved that, using Rahimi-Keshari general method of simulation, a noisy sampling problem as described above with $S(r)U^{(m)}\ket{\alpha}$ as input state requires the "maximum" amount of noise in order to be classically  efficiently simulable.

\section{Input state $U(\chi)S(r)\ket{0}$} \label{input2}
\begin{figure}[t]
\includegraphics[width=0.48\textwidth]{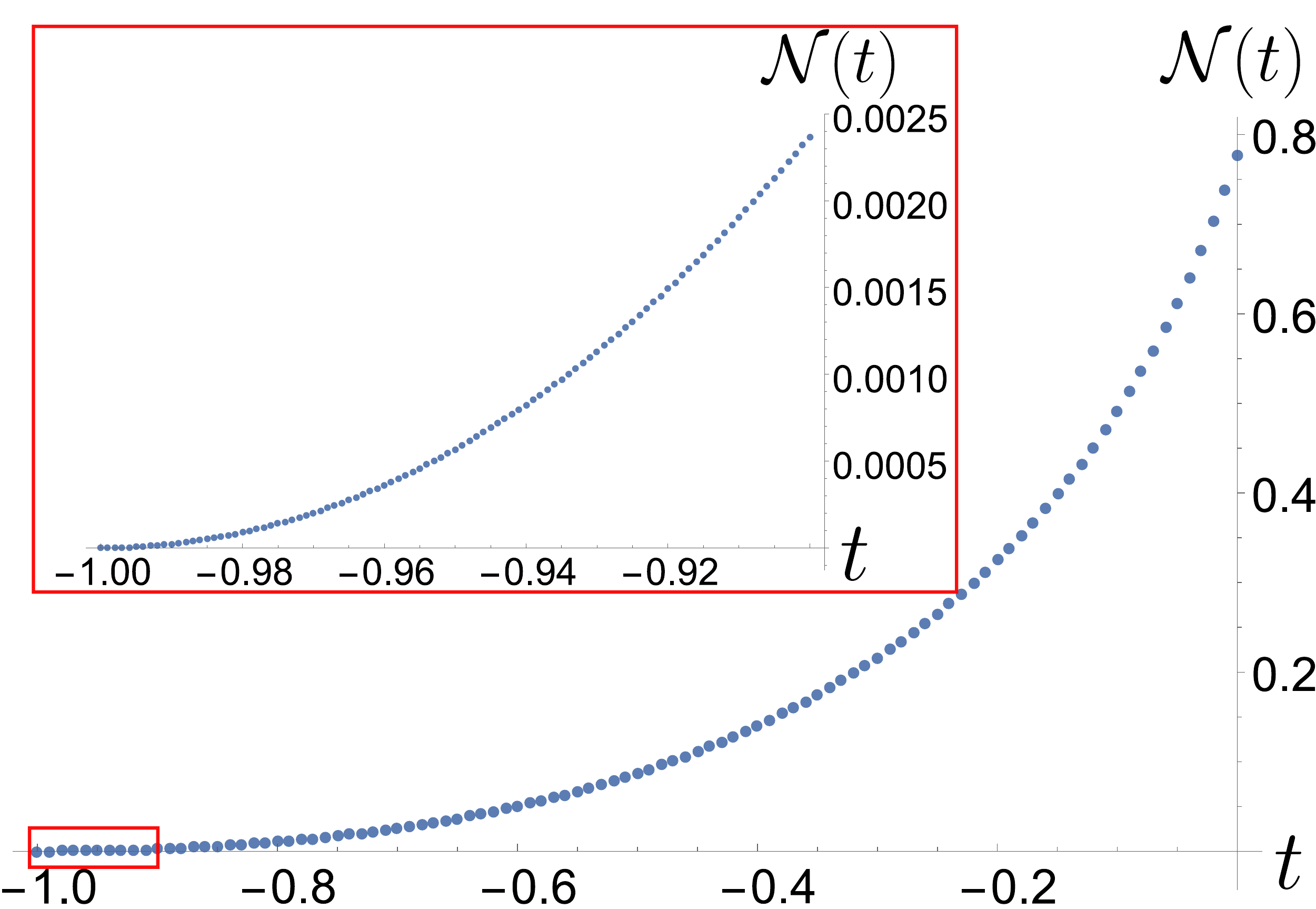}
\caption{\label{input2plot} The plot shows the negativity volume $\mathcal{N}(t)$ of the $t$-PQD associated with the input state $U(\chi=\pi/3)S(r=1)\ket{0}$. 
As can be seen from the inset, the function reaches zero for $t=-1$.
Similar plots and behaviour, i.e. negativity volume approaching zero monotonically for $t=-1$, are obtained $\forall r>0$ and either odd $m>1$ or even $m>4$.}
\end{figure}

We can now focus on another closely related class of initial states, namely $U(\chi)S(r)\ket{0}$. For generic values of the Kerr parameter $\chi$, we once again encounter difficulties in the analytical calculation of the characteristic function in a closed formula.
However, similarly to the previous model, if we consider $\chi=\frac{\pi}{m}$ and apply $U^{(m)}$ to a squeezed vacuum state $S(r)\ket{0}$ we obtain a quantum superposition of squeezed vacuum states \cite{ProdCatKerr}
\begin{equation}
U^{(m)}S(r)\ket{0}=\sum_{q=0}^{m-1}f_q^{(o)}S({r e^{-\frac{4\pi i q}{m}} })\ket{0} \, , \quad m=\text{odd}
\label{superposition_squeezed_vacuum1}
\end{equation}

\begin{equation}
U^{(m)}S(r)\ket{0}=\sum_{q=0}^{m-1}f_q^{(e)}S({r e^{\frac{-4\pi i q+2\pi i}{m}} })\ket{0} \, , \quad m=\text{even}
\label{superposition_squeezed_vacuum2}
\end{equation}
The coefficients $f_q^{(o)}$ and $f_q^{(e)}$ are still given by Eq. (\ref{coeff1}) and Eq. (\ref{coeff2}), respectively.
Recalling how the annihilation operator transforms under the single-mode squeezing unitary operation Eq. (\ref{squeezing})
\begin{equation}
    S^\dagger(re^{i\phi})aS(re^{i\phi})=\mu a + e^{i\phi}\nu a^\dagger \, ,
\end{equation}
where $\mu=\cosh(r)$ and $\nu=\sinh(r)$, the $t$-PQD of Eq. (\ref{superposition_squeezed_vacuum1}) and Eq. (\ref{superposition_squeezed_vacuum2}) is readily obtained once we have the $t$-PQD of $S(re^{i\phi})\ketbra{0}S^\dagger(re^{i\psi})$.
As outlined in Appendix \ref{A}, we find the characteristic function of this dyadic 
\begin{equation}
\begin{split}
    \phi^{(t)}(\xi) & = \Tr{S(re^{i\phi})\ketbra{0}S^\dagger(re^{i\psi}) D(\xi)}e^{\frac{t}{2}\vert\xi\vert^2} 
\end{split}    
\end{equation}
to have the following form
\begin{equation}
    \phi^{(t)}(\xi)=\tilde{\mu}^{-\frac{1}{2}} e^{-\frac{1}{2}\vert \xi\mu-\xi^*\nu e^{i\phi}\vert^2 +\frac{\tilde{\nu}}{2\tilde{\mu}}e^{-i\tilde{\phi}}(\xi\mu-\xi^*\nu e^{i\phi})^2-\frac{i\Phi}{4}+\frac{t}{2}\vert\xi\vert^2} \, .
    \label{final}
\end{equation}
Here $\mu=\cosh(r)$, $\nu=\sinh(r)$, $\tilde{\mu}=\cosh(\tilde{r})$ and  $\tilde{\nu}=\sinh(\tilde{r})$. The remaining parameters $\tilde{r},\tilde{\phi}$ and $\Phi$ are defined by Equations (\ref{modulus}) and (\ref{phase}), respectively.
A Fourier-transform of the characteristic function yields the analytical expression of the $t$-PQD of $S(re^{i\phi})\ketbra{0}S^\dagger(re^{i\psi})$. With this we straightforwardly obtain the $t$-PQD of the initial state, numerically compute its volume of negativity and find the threshold value $\overline{t}$ for which the $t$-PQD is non-negative $\forall t < \overline{t}$.
We once again find strong numerical evidence that the $t$-PQD of $U(\chi=\pi/m)S(r)\ket{0}$ becomes non-negative for $s\leq \overline{t}=-1\quad \forall r>0$ and either odd $m>1$ or even $m>4$.
Figure \ref{input2plot} displays, with a specific example, the typical features of the negativity volume associated with the input state $U(\chi=\pi/m)S(r)\ket{0}$ $t$-PQD. 
The reason for this unusual behaviour is that $U^{(m)}S(r)\ket{0}$ with $m=2$ or $m=4$ are Gaussian states and, as such, their $t$-PQD is non-negative by definition for every value of the ordering parameter $t$ for which the function is well defined and their negativity volume is strictly zero. In particular, one finds that in these two cases $\overline{t}=e^{-2r}$, i.e. the result one obtains for a squeezed vacuum with squeezing parameter $r$, hence the Kerr non-linearity does not provide any advantage in these scenarios.
However, these two ``anomalies'' vanish if we add displacement to our initial state and thus consider $U(\chi)S(r)\ket{\alpha}$, as we will show for $m=2$ shortly.
(Note how this choice would also constitute a fairer comparison to the other state we considered, namely $S(r)U(\chi)\ket{\alpha}$.)
First of all, we show that $U^{(2)}S(r)\ket{0}=S(-r)\ket{0}$. This can easily be seen by expanding the squeezed vacuum on the Fock basis
\begin{equation}
    S(r)\ket{0}=\frac{1}{\sqrt{\cosh{r}}}\sum_{n=0}^{\infty}[\tanh{r}]^n \frac{\sqrt{(2n)!}}{2^n n!} \ket{2n}
\end{equation}
and using $U^{(2)}\ket{2n}=(-1)^n \ket{2n}$.
\\
This in turn means that $U^{(2)}S(r)U^{(2)\dagger}=S(-r)$. With this identity we can write
\begin{equation}
    U^{(2)}S(r)\ket{\alpha}=U^{(2)}S(r) U^{(2)\dagger} U^{(2)}\ket{\alpha}=S(-r)U^{(2)}\ket{\alpha}
\end{equation}
and we immediately realize that this state is just the squeezed cat state we have already discussed in the previous section. Hence, if we add the displacement to the initial state we once again obtain $\overline{t}=-1$ even for $m=2$.
We expect the same to happen $\forall m>1$, given the cat-like structure of $U^{(m)}S(r)\ket{\alpha}$.
\\
\section{Conclusions}
\label{conclusions}
In this work we have investigated the possibility of introducing higher-order non-linearities into the Gaussian Boson sampling framework so as to enhance the computational complexity of the task and 
consequently increase the inefficiencies that allow for a classical simulation to be feasible.
Using a phase-space formalism based on the negativity of the relevant PQDs, we have established a necessary non-classicality test that any experimental demonstration of quantum advantage must satisfy. This sufficient condition for an efficient classical simulation for noisy Boson sampling is formulated in terms of  inequalities that involve the noise parameters characterising the system. 
In this paper we have used noise to gauge how inefficient it is to simulate a given boson sampling task classically.
Our results indeed suggest that the addition of single-mode Kerr non-linearity at the input state preparation level, while retaining a linear-optical evolution, makes the protocol more robust to noise and relaxes the constraints on the noise parameters required to show quantum advantage.
A possible limitation of the formalism we employed is that it only allows us to make predictions about the existence of efficient classical exact simulations. Future efforts might focus on approximate simulation methods of noisy Boson sampling tasks in the presence of non-linear operations as well as studying the role of thermal effects in a general multi-mode setting. 
Another interesting direction for future research include investigating the role of other classes of higher-order non-linearities, different from single-mode Kerr operations, in increasing the computational complexity of Boson sampling problems.

\section{Acknowledgments}
G.B. is part of the AppQInfo MSCA ITN which received funding from the European Union’s Horizon 2020 research and innovation programme under the Marie Sklodowska-Curie grant agreement No 956071.
H.K. is supported by the KIAS Individual Grant No. CG085301 at Korea Institute for Advanced Study.
This work is supported by the KIST Open Research Program. This work is supported by the UK Hub in Quantum Computing and Simulation, part of the UK National Quantum Technologies Programme with funding from UKRI EPSRC grant EP/T001062/1.
MSK thanks Guillaume Thekkadath for discussions.

\appendix

\section{}
\label{A}
Next we outline the techniques employed in the calculation of the characteristic function of $S(re^{i\phi})\ketbra{0}S^\dagger(re^{i\psi})$.
\begin{equation}
\begin{split}
    \phi^{(t)}(\xi) & = \Tr{S(re^{i\phi})\ketbra{0}S^\dagger(re^{i\psi}) D(\xi)}e^{\frac{t}{2}\vert\xi\vert^2} 
   \\ & = \bra{0}S^\dagger(re^{i\psi}) D(\xi)S(re^{i\phi})\ket{0} e^{\frac{t}{2}\vert\xi\vert^2} 
    \\ & = \bra{0}S^\dagger(re^{i\psi})S(re^{i\phi})S^\dagger(re^{i\phi}) D(\xi)S(re^{i\phi})\ket{0} e^{\frac{t}{2}\vert\xi\vert^2} 
\end{split}    
\end{equation}
We can then use $S^\dagger(re^{i\phi}) D(\xi)S(re^{i\phi})= D(\xi\mu-\xi^*\nu e^{i\phi})$.
\\
\\
Before moving on, it is useful to show how to compose two single-mode squeezing operations.
Let us consider a generic single-mode squeezing operation $S(\xi_i)$ with squeezing parameter $\xi_i=r_i e^{i\phi_i}$ and let us define $\zeta_i\doteq\tanh(r_i)e^{i\phi_i}$. One can then prove \cite{agarwal2012quantumoptics} the following identity
\begin{equation}
    S(\xi_1)S(\xi_2)=S(\xi_3)e^{i\Phi(\xi_1,\xi_2)( \frac{a^\dagger a + 1/2}{2} )} \, ,
    \label{squeezing_composition}
\end{equation}
where 
\begin{equation}
    \zeta_3=\frac{\zeta_1+\zeta_2}{1+\zeta_1^* \zeta_2} 
    \label{modulus}
\end{equation}
and 
\begin{equation}
    \Phi(\xi_1,\xi_2)=-i\log \left( \frac{1+\zeta_1 \zeta_2^*}{1+\zeta_1^* \zeta_2}\right) \, .
    \label{phase}
\end{equation}
Recall that $\mathfrak{su}(1,1)$ generators $\lbrace K_+,K_-,K_0 \rbrace$ satisfy the commutation rules \cite{su11algebra}
\begin{equation}
    [K_-,K_+]=2K_0 \, , \quad [K_0,K_\pm]=\pm K_\pm \, .
\end{equation}
The single-mode bosonic representation of this algebra is given by 
\begin{equation}
    K_+ = \frac{a^{\dagger 2}}{2}\, , \quad\quad K_- = \frac{a^{ 2}}{2} \, ,\quad\quad K_0 = \frac{1}{2}\left(a^\dagger a +\frac{1}{2}\right)
\end{equation}
An easy way to verify the squeezing composition rule (\ref{squeezing_composition}) is to use the following matrix representation of $\mathfrak{su}(1,1)$
\begin{equation}
    K_+ = \mqty(0 & 1 \\ 0 & 0)\, ,\quad  K_- = \mqty(0 & 0 \\ -1 & 0)\, , \quad  K_0 = \mqty(1/2 & 0 \\ 0 & -1/2)\, .
\end{equation}
Using the properties of the $\mathfrak{su}(1,1)$ algebra one can also prove the following well known decomposition of the single-mode squeezing operator
\begin{equation}
S(re^{i\phi})=e^{\frac{\nu e^{i\phi}}{2\mu}a^{\dagger 2}} \mu^{-a^ \dagger a - 1/2} e^{-\frac{\nu e^{-i\phi}}{2\mu}a^2}
\label{squeezing_decomposition}
\end{equation}
Hence, using Eq. (\ref{squeezing_composition}) we can write
\begin{equation}
    S(-re^{i\phi})S(re^{i\psi}) = S(\tilde{r}e^{i\tilde{\phi}})e^{i\Phi( \frac{a^\dagger a + 1/2}{2} )} \, ,
\end{equation}
where $\tilde{r},\tilde{\phi}$ and $\Phi$ are defined by Equations (\ref{modulus}) and (\ref{phase}), respectively.
Using Eq. (\ref{squeezing_decomposition}) one then finally obtains the characteristic function of $S(re^{i\phi})\ketbra{0}S^\dagger(re^{i\psi})$ displayed in Eq. (\ref{final}) of the main text.

\section{}
\label{B}
In order to take thermal effects into account we consider a modification of the loss model described in the main text, where each of the $M$ additional environmental modes are now in a thermal state. It then follows that the action of the lossy LON is now described by the map 
\begin{equation}
    \mathcal{E}^\prime(\ketbra{\bm{\gamma}})=\Tr{\mathcal{U} \ketbra{\bm{\gamma}} \otimes \nu_{th}^{\otimes M} \mathcal{U}^\dagger} \, ,
\end{equation}
where $\mathcal{U}$ is, once again, the unitary operator associated with the larger $2M$-mode interferometer.
$\nu_{th}$ represents a thermal state,  i.e.,
\begin{equation}
    \nu_{th}= \frac{1}{1+\overline{n}} \left(\frac{\overline{n}}{1+\overline{n}}\right)^{a^\dagger a} \, ,
\end{equation}
where $\overline{n}$ is the mean number of photons and $a,a^\dagger$ are the  annihilation operator and creation operator of the mode, respectively.
We remind the reader that the action of the quantum channel $\mathcal{E}$ on a $M$-mode coherent state is all we need to compute the transition function $T_{\mathcal{E}}^{(\bm{s},\bm{t})}$ and that the latter is independent of the input states and the final measurements.
\\
In order to make the calculations easier, let us consider the single-mode $M=1$ case, i.e. a toy model where the lossy LON is just a beam splitter, characterized by transmittivity $(\cos{\theta})^2=\eta_L$, that couples a coherent state $\ket{\gamma}$ with a thermal state $\nu_{th}(k)$, and we then trace over the environmental degrees of freedom. Here $k>1$ is the value of the quadrature variances of the thermal state and can also be expressed as $k=2\overline{n}+1$, where $\overline{n}$ is the mean number of photons. 
\\
First, it is useful to see what happens in the zero temperature case, i.e. we specialize Eq. (\ref{coh_state_trans}) for a single mode and obtain
\begin{equation}
    \mathcal{E}(\ketbra{\gamma})=\ketbra{\gamma \cos{\theta}}\equiv\ketbra{\tilde{\gamma}} \, ,
\end{equation}
where we defined $\tilde{\gamma}=\gamma\cos{\theta}$.
Hence, in this scenario, the transfer matrix is simply a real number $\bm{L}=\cos{\theta}=\sqrt{\eta_L}$.
\\
Moving onto the finite temperature case, using the Gaussian formalism one easily shows that
\begin{equation}
\begin{split}
    \mathcal{E}^\prime(\ketbra{\gamma})& =
    \Tr_{env}\lbrace {\mathcal{U}(\theta) (\ketbra{\gamma} \otimes \nu_{th})\, \mathcal{U}^\dagger(\theta)} \rbrace \\ & =D(\tilde{\gamma})\nu_{th}(\lambda)D^\dagger(\tilde{\gamma})
\end{split}
\end{equation}
where $\mathcal{U}(\theta)$ is now the beam splitter unitary operator and $\lambda=(\cos{\theta})^2+k (\sin{\theta})^2$ is a real parameter strictly greater than 1 and the trace is taken over the environmental degrees of freedom.
We can then compute
\begin{equation}
\begin{split}
    \Tr{\mathcal{E}^\prime(\ketbra{\gamma})D(\zeta)} & =
    \Tr{D(\tilde{\gamma})\nu_{th}(\lambda)D^\dagger(\tilde{\gamma})D(\zeta)} \\ & = 
    e^{\zeta\tilde{\gamma}^*-\zeta^*\tilde{\gamma}}\Tr{\nu_{th}(\lambda)D(\zeta)} \, .
\end{split}
\end{equation}
The trace in the last expression is evaluated by exploiting the P-function representation of the thermal state, i.e.
\begin{equation}
    \begin{split}
        \Tr{\mathcal{E}^\prime(\ketbra{\gamma})D(\zeta)} & = \Tr{\int d^2\beta P(\beta)\ketbra{\beta}D(\zeta)} \\ & = e^{-\frac{\lambda}{2}(\zeta_1^2+\zeta_2^2)} \, ,
    \end{split}
    \label{passaggio1}
\end{equation}
where
\begin{equation}
    P(\beta)=\frac{2}{\pi(\lambda-1)}e^{-\frac{2}{\lambda-1}(\beta_1^2+\beta_2^2)}
\end{equation}
is the P-function of $\nu_{th}(\lambda)$.
Now plugging Eq. (\ref{passaggio1}) into Eq. (\ref{expansion}) and using the identity
\begin{equation}
    \int d^2\beta e^{\zeta\beta^*-\zeta^*\beta}=\pi^2\delta^{(2)}(\zeta)
\end{equation}
we obtain 
\begin{equation}
    \Tr{\mathcal{D^\dagger(\xi)}D(\zeta)}=\pi\delta^{(2)}(\xi-\zeta\cos{\theta})e^{\frac{\vert\zeta\vert^2}{2}(\cos^2{\theta}-\lambda)} \, .
\end{equation}
Substituting this last expression into Eq.(\ref{transition}) yields the transition function $T_{\mathcal{E}^\prime}^{(s,t)}$.
\begin{equation}
    \begin{split}
        T_{\mathcal{E}^\prime}^{(s,t)}(\alpha,\beta)=  \int \frac{d^2\zeta}{\pi^2} &  e^{-\frac{\vert\zeta\vert^2}{2}(t\cos^2{\theta} -s+\lambda-\cos{\theta})} \cdot
        \\ & e^{\zeta(\alpha^*\cos{\theta}-\beta^*)-\zeta^*(\alpha\cos{\theta}-\beta)} \, .
    \end{split}
\end{equation}
Hence, the function is well-behaved and has Gaussian form as long as 
\begin{equation}
    t(\cos{\theta})^2-s+\lambda-(\cos{\theta})^2\geq 0 \, .
\end{equation}
On the other hand, inequality Eq. (\ref{trans_condition}) for a single mode reads
\begin{equation}
    t(\cos{\theta})^2-s+1-(\cos{\theta})^2\geq 0 \, .
\end{equation} 
Hence, we have obtained a very similar inequality, where thermal effects are entirely accounted for in the parameter $\lambda>1$. 
Note that the zero-temperature expression is retrieved for $\lambda=1$. We can finally 
use the technique outlined in the main text to compute the noise thresholds that allow for an efficient simulation of the sampling task on a classical machine. In particular, one finds
\begin{equation}
    \frac{p_D}{\eta_D}\geq\eta_L + \frac{1-\lambda}{2} \, .
\end{equation}
If we then express $\lambda$ in terms of $\overline{n}$ and $\eta_L$ we obtain 
\begin{equation}
     \frac{p_D}{\eta_D}\geq\eta_L - \overline{n}(1-\eta_L) \, .
     \label{final_ineq}
\end{equation}
The term $\overline{n}(1-\eta_L)$ represents the correction to the results we presented in main text (Eq. (\ref{result})), for the $M=1$ case, when temperature effects are taken into account.
As expected, the additional thermal noise has the effect of \textit{reducing} the noise in the detection which is sufficient to efficiently simulate the task on a classical machine. 
We also notice that if 
\begin{equation}
    \overline{n}\geq \frac{\eta_L}{1-\eta_L}
\end{equation}
then the right hand side of the inequality Eq. (\ref{final_ineq}) becomes negative and the sampling problem becomes classically simulable even with ideal detectors. This is indeed expected, as we know that Boson sampling with thermal state inputs - or any other classical input state - is efficiently simulable. As a result, we envision a transition in the computational complexity of the problem as the temperature of the environment grows.
\section{}
\label{C}
Here we provide the proof to obtain the identity Eq. (\ref{kerr_transformation}). We exploit the Baker-Campbell-Hausdorff formula
\begin{equation}
    e^{A}Be^{-A} = B+[A,B]+\frac{1}{2!}[A,[A,B]] + \dots \, ,
\end{equation}
with the following substitutions
\begin{equation}
    A = i\chi a^{\dagger 2} a^2 \quad\quad B=a \, .
\end{equation}
One then obtains
\begin{equation}
\begin{split}
     U^\dagger(\chi) a U(\chi) & = a -2i\chi (a^\dagger a)a + \frac{(-2i\chi)^2}{2!} (a^\dagger a)^2 a + \dots
     \\ & = \sum_{n=0}^\infty  \frac{(-2i\chi a^\dagger a)^n}{n!} a = e^{-2i\chi a^\dagger a}a \, .
\end{split}
\end{equation}
\bibliography{biblio}

\end{document}